\documentstyle[12pt,aaspp4]{article}
\begin{document}
\parindent=1.0cm

\title
{THE UPPER ASYMPTOTIC GIANT BRANCH OF THE ELLIPTICAL GALAXY MAFFEI 1, AND 
COMPARISONS WITH M32 AND NGC 5128}

\author{T. J. Davidge \altaffilmark{1}$^{,}$ \altaffilmark{2}}

\affil{Canadian Gemini Office, Herzberg Institute of Astrophysics,
\\National Research Council of Canada, 5071 W. Saanich Road, \\ Victoria,
British Columbia, Canada V9E 2E7 \\ {\it email: tim.davidge@nrc.ca}}

\altaffiltext{1}{Visiting Astronomer, Canada-France-Hawaii Telescope, which is
operated by the National Research Council of Canada, the Centre National de le
Recherche Scientifique, and the University of Hawaii.}

\altaffiltext{2}{Visiting Astronomer, Gemini Observatory, which is
operated by the Association of Universities for Research in Astronomy, Inc., 
under a cooperative agreement with the NSF on behalf of the Gemini 
partnership: the National Research Council (Canada), the National Science 
Foundation (United States), the Particle Physics and Astronomy Research 
Council (United Kingdom), the Australian Reasearch Council (Australia), 
CNPq (Brazil), and CONICET (Argentina).}

\begin{abstract}

	Deep near-infrared images obtained with adaptive optics (AO) systems on 
the Gemini North and Canada-France-Hawaii telescopes are used to investigate 
the bright stellar content and central regions of the nearby elliptical 
galaxy Maffei 1. Stars evolving on the upper asymptotic giant branch (AGB) are 
resolved in a field 3 arcmin from the center of the galaxy. The locus of bright 
giants on the $(K, H-K)$ color-magnitude diagram is consistent with a 
population of stars like those in Baade's Window reddened by $E(H-K) = 0.28 \pm 
0.05$ mag. This corresponds to A$_V = 4.5 \pm 0.8$ mag, and is consistent with 
previous estimates of the line of sight extinction computed from the integrated 
properties of Maffei 1. The AGB-tip occurs at $K = 20.0$, which 
correponds to M$_K = -8.7$; hence, the AGB-tip brightness in 
Maffei 1 is comparable to that in M32, NGC 5128, and the bulges of M31 and the 
Milky-Way. The near-infrared luminosity functions (LFs) of bright AGB stars in 
Maffei 1, M32, and NGC 5128 are also in excellent agreement, both in terms of 
overall shape and the relative density of infrared-bright 
stars with respect to the fainter stars that 
dominate the light at visible and red wavelengths. It is 
concluded that the brightest AGB stars in Maffei 1, NGC 5128, M32, and the 
bulge of M31 trace an old, metal-rich population, rather than an intermediate 
age population. It is also demonstrated that Maffei 1 contains a distinct 
red nucleus, and this is likely the optical signature of low-level nuclear 
activity and/or a distinct central stellar population. Finally, there is an 
absence of globular clusters brighter than the peak of the globular cluster LF
in the central $700 \times 700$ parsecs of Maffei 1.

\end{abstract}


\section{INTRODUCTION}

	The brightnesses and spatial distributions of stars 
evolving on the asymptotic giant branch (AGB) provide clues about the past 
evolution of galaxies. Although stars near the AGB-tip are relatively bright, 
efforts to resolve these objects in the dense main bodies of nearby spheroids 
typically require angular resolutions approaching the diffraction limit 
of 2.5-metre or larger telescopes. Spheroids within the Local Group are obvious 
first targets for any studies of resolved stellar content, and 
Davidge (2000a), Davidge et al. (2000), and Davidge (2001a) found that the 
brightest AGB stars in the compact elliptical galaxy M32 and the 
bulge of M31 have similar brightnesses, and are well mixed with 
the fainter stars in these systems. This suggests that the 
most luminous AGB stars in M32 and the bulge of M31 belong to a population that
formed when the structural characteristics of these galaxies were imprinted 
(Davidge 2001a). A similar AGB population may be present in the bulge of 
the Milky-Way (Davidge 2001a) -- a system that deep photometric studies 
indicate has an old age (Feltzing \& Gilmore 2000; Ortolani et al. 1995). 

	Davidge (2001a) suggested that the bright AGB stars detected in Local 
Group spheroids trace an old, metal-rich population, and this suggestion can be 
tested by examining the brightness and spatial distribution of AGB stars 
in a larger sample of spheroids. With a distance between 4 and 4.5 Mpc (Davidge 
\& van den Bergh 2001; Luppino \& Tonry 1993), Maffei 1 is one of the closest 
large elliptical galaxies, and hence is an obvious target for efforts to study 
the bright stellar content. A complicating factor is that 
Maffei 1 is viewed through the Galactic disk, and so is subject 
to significant foreground extinction. Buta \& McCall (1983) concluded that 
A$_V = 5.1 \pm 0.2$ based on the integrated color of Maffei 1 and the column 
density of hydrogen along the line of sight. Hence, efforts to resolve 
stars in Maffei 1 will likely be most successful in the near-infrared, which 
is also the prime wavelength regime for investigating stars on the upper AGB.
Davidge \& van den Bergh (2001; hereafter DvdB) investigated the near-infrared 
photometric properties of the brightest AGB stars in a field with a projected 
distance of 6 arcmin from the center of Maffei 1. Despite 
long integration times (3 hours per filter) and image quality 
close to the theoretical diffraction limit of the CFHT, only 
stars within $\sim 1.5$ mag of the AGB-tip were detected. 

	In the present study, two datasets are used to 
investigate different regions of Maffei 1. The first dataset 
consists of deep $H$ and $K'$ images obtained with the University of Hawaii 
Adaptive Optics (UHAO) system on the Gemini North telescope, which sample a 
field 3 arcmin from the center of Maffei 1. 
Not only is the stellar density higher than in the field studied by DvdB, but 
stars that are 1 mag in $K$ fainter than in the DvdB dataset are detected, 
permitting a detailed comparison with the bright stellar contents of other 
galaxies.

	The second dataset consists of $J, H,$ and $Ks$ images of the center of 
Maffei 1 that were obtained with the CFHT AO system. While not 
diffraction-limited, these data have a resolution that permits 
the central $700 \times 700$ parsecs of the galaxy to be investigated at 
sub-arcsec angular scales. Data of this nature can determine if Maffei 1 has a 
photometrically distinct nucleus, which in turn provides clues about the past 
evolution of this galaxy.

	The paper is structured as follows. Details of the observations and 
the data reduction techniques are presented in \S 2, while the photometric 
properties of stars in the Gemini dataset are discussed and compared with those 
of other galaxies in \S 3 and 4. The photometric properties of the central 
regions of Maffei 1 are examined in \S 5, while the results of a search for 
globular clusters in both datasets is presented in \S 6. A summary and 
discussion of the results of this paper follows in \S 7.

\section{OBSERVATIONS \& REDUCTIONS}

	$H$ and $K'$ images of a field located $3\farcm2$ from the 
center of Maffei 1, which will be refered to as the `deep' 
field throughout this paper, were recorded during the night 
of UT September 16 2001 with the UHAO system  
and QUIRC imager, which were mounted at the f/16 Cassegrain focus of the 8 
metre Gemini North telescope. The detector in QUIRC is a $1024 \times 1024$ 
Hg:Cd:Te array, with each pixel subtending $0\farcs02$ on a side, so that a 
$20\farcs5 \times 20\farcs5$ field is imaged. A detailed description of the 
UHAO system has been given by Graves et al. (1998).

	The UHAO system uses natural guide stars as reference sources to 
monitor wavefront distortion, and the guide star for these observations was the 
$R = 13.4$ star GSC 03699-01165 ($\alpha$ = 02:36:11, $\delta$ = $+$59:40:15 
E2000). Data in $K'$ were recorded at three pointings, with the guide star 
offset by $3\farcs5$ between each of these. While this offset is significant 
when compared with the QUIRC science field, and hence has a major impact on 
the angular extent of the final field, it was employed so that 
a calibration frame for removing interference fringes 
and thermal emission signatures could be constructed from these data alone. 
$5 \times 60$ sec exposures were recorded 
at the corners of a $2\arcsec \times 2\arcsec$ square dither pattern 
for each pointing, and so the total exposure time in $K'$ is 3 pointings 
$\times 4$ dither positions per pointing $\times$ 5 exposures per dither 
position $\times 60$ sec per exposure $= 3600$ seconds. Images in $H$ were 
recorded using the same dither pattern and 60 sec exposure time, 
although data were recorded at only 2 pointings due to time restrictions, so 
that the total integration time in this filter is 
2400 seconds (i.e. two thirds that in $K'$).

	The final images have FWHM = $0\farcs13$ ($H$) and $0\farcs14$ ($K'$), 
which are larger than the $\frac{\lambda}{D} = 0\farcs04 (H)$ and $0\farcs06 
(K')$ telescope diffraction limits. These data thus have low Strehl ratios, 
due mainly to the seeing being slightly worse than the Mauna Kea 
median on this night, with an uncorrected FWHM $\sim 1\farcs0$ at 
visible wavelengths. The UHAO uses a 36 element wavefront sensor; while 
this is adequate to deliver diffraction-limited images at near-infrared 
wavelengths on a 4 metre telescope during median Mauna Kea seeing conditions, 
good seeing is required to achieve a high Strehl 
ratio with this order of compensation on an 8 metre telescope.
Despite having low Strehl ratios, the angular resolution of these 
data is slightly better than was obtained by DvdB, 
who achieved FWHM $= 0\farcs14$ in $H$ and $0\farcs15$ in $Ks$. 

	Two standards from Hawarden et al. (2001) were observed 
on the same night that the Maffei 1 data were recorded. 
A growth curve analysis established that a $1\farcs5$ aperture contained, for 
all practical purposes, all of the light from these stars. 
The zeropoints predicted from the two standards 
differ by only 0.06 mag in $K$, and the mean zeropoint agrees to 
within 0.1 mag of that measured during previous UHAO $+$ QUIRC observing runs 
(e.g. Davidge et al. 2000).

	The center of Maffei 1 was observed with the CFHT AOB and KIR imager 
during the night of UT September 10 2000. The detector in KIR is a $1024 \times 
1024$ Hg:Cd:Te array, with each pixel subtending $0\farcs034$ on a side, 
so that the total imaged field is $34\farcs8 \times 34\farcs8$. 
The CFHT AO system, which has been described by 
Rigaut et al. (1998), uses natural guide stars to monitor wavefront 
distortion, and the $R = 12.3$ star GSC 03699-01147, which is 
$0\farcm43$ from the center of Maffei 1, was used as the reference source for 
these observations. A single 60 sec exposure was recorded at each 
corner of a $1\arcsec \times 1\arcsec$ square dither pattern through 
$J, H,$ and $Ks$ filters; the total integration time is thus $4 \times 60 = 
240$ sec per filter.

	Stars in the final CFHT images have FWHM $= 0\farcs25$ in $J$, 
$0\farcs30$ in $H$, and $0\farcs35$ in $Ks$. The angular resolution is thus 
much larger than the theoretical diffraction limits of the CFHT, which 
are $\frac{\lambda}{D} = 0\farcs07, 0\farcs09,$ and $0\farcs12$ in $J$, $H$, 
and $Ks$. The low Strehl ratios of these data is due to poor natural 
seeing conditions when they were recorded. Nevertheless, these data 
still have an angular resolution that is superior to any previous 
observations of the center of Maffei 1 at these wavelengths.

	Six standard stars were observed with the CFHT AOB $+$ KIR on the 
nights of Sept 10 and 11 2000. The standard deviations in the photometric 
zeropoints are $\pm 0.03, \pm 0.02,$ and $\pm 0.04$ mag in $J$, $H$, and $Ks$. 
The photometric zeropoints computed from these data are 
consistent those derived on previous runs using the same instrumental 
configuration (e.g. Davidge \& Courteau 1999a; Davidge et al. 2000).

	The data reduction sequence for the raw images from both the CFHT and 
Gemini datasets was as follows: (1) dark subtraction, (2) division by a dome 
flat, (3) subtraction of a calibration frame to remove thermal 
artifacts and interference fringes, and (4) subtraction 
of the DC sky level at each dither position. The processed images in each 
filter were then registered, median-combined, and trimmed to 
the area having full integration time; the final images of the deep field 
thus cover $14\farcs6 \times 14\farcs4$, while those of the central 
field cover $34\farcs0 \times 34\farcs0$. The final $K'$ and $Ks$ 
images of the deep and central fields are shown in Figures 1 and 2. 
Individual stars in Maffei 1 are clearly evident in Figure 1.

\section{THE MAFFEI 1 DEEP FIELD}

\subsection{Photometry of Stars in the Maffei 1 Deep Field}

	The brightnesses of stars in the deep field were measured with the 
PSF-fitting program ALLSTAR (Stetson \& Harris 1988). Target lists, 
preliminary brightnesses, and PSFs were obtained using DAOPHOT (Stetson 
1987) tasks. There are indications that the PSF is very uniform across the 
field, and this is demonstrated in Figure 3, where the mean $K'$ light profiles 
of stars in two radial intervals, centered on the guide star and selected to 
sample roughly equal areas, are compared. The comparison in Figure 3 should be 
viewed with some caution, as FWHM is not an unambiguous estimator of Strehl 
ratio, while information about any radial elongation of stellar 
images in the direction of the guide source, which is a classical signature 
of anisoplanicity (e.g McClure et al. 1991), is largely
lost when considering radially-averaged profiles of the type 
shown in this figure. Nevertheless, based on the uniformity evident in Figure 
3, it was decided to construct a single PSF for each filter. 
The absence of obvious signatures of anisoplanicity is 
not surprising given previous experience with the CFHT AO system, which 
indicates that the PSF is stable over the $34\farcs8 \times 34\farcs8$ KIR 
science field for moderate orders of AO compensation during typical atmospheric 
conditions (e.g. Davidge \& Courteau 1999a; Davidge 2001b). Not only does the 
Maffei 1 deep field, with dimensions of $14\farcs6 \times 14\farcs4$, 
cover a smaller area than the CFHT KIR field but, when used 
on an 8 metre telescope, the UHAO, which has a 36 element wave front sensor 
(WFS), delivers a lower order of correction than the 19 element WFS AO system 
used on the 3.6 meter CFHT, and this further contributes to PSF uniformity. 

	The photometric calibration was defined using the standard star 
observations discussed in \S 2, with the instrumental $K'$ measurements 
being transformed into $K$ magnitudes. Wainscoat \& Cowie (1992) find that 
there is a significant color term between $K'-K$ and $H-K$. As only two 
standards were observed to calibrate the Maffei 1 deep field data, a relation 
between $K'-K$ and $H-K$ was derived from the entries in Table 1 of 
Wainscoat \& Cowie (1992). While those authors chose to fit a 
linear relation to their data over the full range of colors, the entries in 
their Table 1 suggest that there is a break in the relation near $H-K = 0.4$, 
in the sense that $K-K' \sim 0.06$ when $H-K \geq 
0.4$, and so the calibration of the deep field data assumes that $K-K' = 
0.06$ for very red colors. The uncertainty in the transformation equation 
color term likely contributes $\pm 0.05$ mag uncertainty to the Maffei 1 
photometry.

	The order of AO correction depends on factors such as atmospheric 
conditions and the brightness of the AO guide star, and for these reasons the 
standard stars likely have different Strehl ratios than the stars in the Maffei 
1 deep field. To avoid complications introduced by such differences in Strehl 
ratio, the Maffei 1 data were calibrated using large radius aperture 
photometry of the PSF stars. All detected neighboring stars were removed 
from the deep field images prior to measuring the brightnesses of the PSF 
stars, and the aperture size was set at the same large radius as the standard 
star measurements. This same calibration procedure has been 
used in previous studies with the CFHT AO system, and it has been demonstrated 
to produce consistent brightnesses from data obtained during different runs 
(Davidge \& Courteau 2002), and with different guide stars (Davidge 2001b). 
Hence, the calibration procedure is insensitive to differences in 
Strehl ratio. This robust calibration procedure was also employed for 
studies of the brightest stars in M32 (Davidge 2000; Davidge et al, 2000) and 
the bulge of M31 (Davidge 2001a) using the same instruments. The observations 
of Maffei 1, M32, and the bulge of M31 thus form a homogeneous dataset for 
comparing the bright stellar content of these systems.

	Artificial star experiments, in which scaled versions of the PSF 
with noise were added to the images using the DAOPHOT ADDSTAR task, were run 
to assess completeness and the uncertainties in the 
photometric measurements due to crowding and photon noise. The 
artificial stars were assigned $H-K = 0.7$ to match the Maffei 1 stellar 
locus (\S 3.2) and $K > 19.5$. The artificial star experiments indicate 
that completeness drops from 100\% at $K = 20$ to 50\% at $K = 22$.

\subsection{The $(K, H-K)$ CMD and $K$ LF of Stars in the Maffei 1 Deep Field}

	The $(K, H-K)$ CMD of the deep field is plotted in Figure 4, where the 
errorbars show the uncertainties predicted from the artificial star 
experiments. Giants in Maffei 1 form a broad plume when $K > 20$. The 
errorbars predicted by the artificial star experiments match the observed 
scatter in $H-K$ in the Maffei 1 giant branch, indicating that the spread in 
the data is due to photometric errors, rather than an intrinsic dispersion in 
stellar properties.

	The peak brightness of the Maffei 1 CMD in Figure 4, which we identify 
as the AGB-tip, occurs at $K = 20$, and this is in excellent agreement with 
what was predicted by DvdB \footnote[3]{The low stellar density in the DvdB 
field, coupled with the modest science field of the KIR imager, resulted in 
a low probability of detecting AGB-tip stars. 
Based on the number density of objects at fainter magnitudes and star counts in 
M32 and the bulge of M31, DvdB predicted that the AGB-tip in Maffei 1 should 
occur near $K=20.0 \pm 0.25$}. The mean color of stars with $K$ between 20.0 
and 21.5 in Figure 4 was computed using an iterative $2-\sigma$ 
rejection threshold to suppress outliers, and the result is $\overline{H-K} = 
0.71 \pm 0.02$, where the uncertainty is the formal error in the mean. 
This mean color is insensitive to the magnitude range from which it is 
computed, and is consistent with the location of the Maffei 1 sequence in the 
$(K, H-K)$ CMD shown in Figure 1 of DvdB. 

	The reddening towards the Maffei 1 deep field can be estimated if it is 
assumed that the brightest stars have the same intrinsic colors as M 
giants in Baade's Window (BW), the brightest of which have $(H-K)_0 \sim 0.43$ 
(Frogel \& Whitford 1987). Since $\overline{H-K} = 0.71$, 
then the color excess of the Maffei 1 deep field is $E(H-K) \sim 0.28$. 
The dominant source of uncertainty in $E(H-K)$ is
the photometric calibration, which has an estimated error of $\pm 0.05$. 
The corresponding extinctions in $K$ and $V$, computed 
using the Rieke \& Lebofsky (1985) reddening curve, are A$_K \sim 0.5 
\pm 0.1$ and A$_V \sim 4.5 \pm 0.8$. The extinction measured from the 
deep field CMD is thus not significantly different from that estimated by 
Buta \& McCall (1983) using other techniques. 

	The adopted reddening affects the true 
distance modulus. Luppino \& Tonry (1993) assumed 
A$_K = 0.63$ to get $\mu_0 = 28.1 \pm 0.25$ from surface brightness 
flucuations, whereas if A$_K = 0.5 \pm 0.1$, based on the color of stars in 
Maffei 1, then $\mu_0 = 28.2 \pm 0.25$. The latter 
distance modulus will be used for the remainder of the paper.

	The $K$ LF of the deep field is shown in Figure 5. The 
completeness-corrected LF follows a power-law, 
with an apparent discontinuity between $K = 20$ and 
$K = 20.5$. The power-law exponent $x = \frac{dlog(n_{0.5})}{dK}$, 
computed from a least squares fit to the LF between $K = 20.25$ and 22.75, is 
$0.70 \pm 0.07$. The LF is therefore significantly different from that of 
first ascent giants in various bulge fields, where $x \sim 0.34$ (Davidge 
2000b). The stars detected in Maffei 1 are intrinsically bright, and 
evolving on the AGB. Stars in the Galactic Bulge with the same intrinsic 
brightness as those in the Maffei 1 deep field have $K_0 < 8.1$; not only 
are such objects relatively rare, but they are usually saturated in imaging 
surveys of the Galactic Bulge. DePoy et al. (1993) examined the $K$ LF of 
bright stars in BW. Their Figure 3 includes data not only from their 
survey, which suffers from saturation effects at the bright end, but 
also from Frogel \& Whitford (1987) and Davidge (1991), and it is 
evident from the combined datasets in this figure that the $K$ LF of 
BW appears to steepen (i.e. $x$ will be significantly larger than the 
value measured from RGB stars) when $K_0 < 8$.

	The RGB-tip in old solar metallicity populations occurs near 
M$_{K}^{RGBT} = -7$ (Bertelli et al. 1994), which corresponds to $K = 21.7$ in 
Maffei 1; therefore, the discontinuity at the bright end in Figure 5 is not due 
to the onset of the RGB. This being said, there is not a clear discontinuity 
near $K = 21.7$ in Figure 5. The absence of an RGB-tip feature in the LF is 
due in part to the metallicity-sensitive nature of M$_{K}^{RGBT}$. The 
stars in the deep field likely span a range of metallicities, 
and this will blur any discontinuity in the LF 
caused by the onset of the RGB. In addition, there are 
significant photometric errors near the faint limit of our data, and these 
further smear any signature of the RGB-tip.

\section{COMPARISONS WITH THE DvdB FIELD, M32, AND NGC 5128}

	In this Section the bright AGB content of the Maffei 1 deep field is 
compared with the stellar contents in the outer regions of Maffei 1, M32, and 
NGC 5128 (Cen A). The latter is the closest 
large elliptical (e.g. Israel 1998, and references therein). These 
comparisons are based on the peak AGB brightness, the shape of the LF, and the 
number density of bright AGB stars, normalized using published surface 
brightness measurements.

	The $K$ LFs of the deep and DvdB fields are compared 
in the upper panel of Figure 6. These fields sample areas of Maffei 1 
that have different stellar densities, and the LF of the DvdB field in 
Figure 6 was scaled along the vertical axis to match the stellar density in 
the deep field according to the $I-$band surface brightness profile of Buta \& 
McCall (1999). 

	The LFs in the upper panel of Figure 6 agree within the estimated 
uncertainties at $K = 21.5$, which is near the expected onset of the RGB 
(\S 3). However, the $K$ LF of the DvdB field falls 
below that of the deep field along the upper 
AGB; the difference between the two LFs at $K = 21$ is significant at 
less than the $2-\sigma$ level, but at $K = 20.5$ the difference is 
significant at almost the $3-\sigma$ level. The comparison in Figure 6
thus suggests that the DvdB field may be deficient in the brightest evolved 
stars when compared with the deep field.

	After accounting for differences in distance and stellar density, 
the bright stellar content in the Maffei 1 deep field is 
representative of that in other nearby elliptical galaxies. 
In the lower panel of Figure 6 the $K$ LFs of the Maffei 1 deep field 
and the M32 outer field studied by Davidge (2000a) are compared. 
We assumed that M32 is equidistant with M31, for which 
a distance modulus $\mu_0 = 24.4$ was adopted (van den Bergh 2000), while 
$\mu_0 = 28.2$ and A$_K = 0.5$ were used for Maffei 1 (see above). The M32 
LF was shifted along the vertical axis so that the stellar density 
matched that in the deep field based on the $V$-band surface brightness 
profile of Maffei 1 from Buta \& McCall (1999) and the $r$-band surface 
brightness profile of M32 from Kent (1987); a correction was also applied 
for the difference in distance. The integrated color of Maffei 1 was assumed to 
be $V - r = 0.25$, which is typical for an elliptical galaxy (Frei \& Gunn 
1994).

	While the M32 LF in Figure 6 falls consistently above the Maffei 1 LF, 
the mean difference is within the systematic uncertainties in the 
relative distances and reddenings of the two systems. Moreover, the peak 
brightnesses in these systems are similar; in M32 the AGB-tip occurs near 
M$_K = -8.9 \pm 0.1$, while in Maffei 1 M$_K = 20 - (28.2 + 0.5) = -8.7 \pm 
0.1$. Thus, the bright stellar contents of the outer 
regions of M32 and the Maffei 1 deep field are not significantly different.

	The M32 outer field studied by Davidge (2000a) samples a low density 
region of this galaxy where crowding is not an issue; hence, the M32 LF in 
Figure 6 can be used to estimate the number of blends in the Maffei 1 data if 
it is assumed that the outer regions of M32 and 
the Maffei 1 deep field have similar stellar contents. When 
shifted to match the distance and surface brightness 
of the Maffei 1 deep field, the M32 LF predicts that 
there should be 3.2 stars per 0.5 mag interval per arcsec$^{2}$ 
at $K = 22.5$. If each resolution element in the Maffei 1 data has 
a radius that is one half the FWHM, then this corresponds to 0.05 stars per 
0.5 mag interval per resolution element at $K = 22.5$. The probability 
of two stars with $K = 22.5$ falling in the same resolution element, and 
thereby creating a blended object with $K \sim 22$, is then 0.2\%. This simple 
calculation shows that blending is not an issue in the Maffei 1 deep field 
dataset. 

	NGC 5128 is an interesting comparison object for Maffei 1 as 
these galaxies have roughly similar distances (3.5 Mpc for NGC 5128 
versus 4.4 Mpc for Maffei 1) and integrated brightnesses (M$_V \sim 
-22$); moreover, neither galaxy is in a dense environment. 
Marleau et al. (2000) discuss F160W NIC2 observations of a field 
with a projected distance of 9 kpc from the center of NGC 5128, and in Figure 7
the $H$ LFs from their data and the Maffei 1 deep field are compared; the 
population of bright foreground stars with [F110W] -- [F160W] $\sim 0.8$ in the 
Marleau et al. (2000) dataset were not included in the NGC 5128 LF.
The NGC 5128 data shown in this figure were shifted by 
1.2 mag to account for the greater apparent distance of 
Maffei 1 in $H$ assuming that $\mu_0 = 27.75$ for NGC 5128 
(Marleau et al. 2000). The resulting LF was then scaled to match the 
stellar density in the Maffei 1 field based on the van den Bergh (1976) 
$V-$band surface brightness profile after correcting for 
the differences in distance. 

	There is excellent agreement between the $H$ LFs of NGC 5128 and 
Maffei 1 in Figure 7. A potential concern is 
that the brightest stars in Maffei 1 have $H = 
20.7$, whereas the brightest stars in the Marleau et al. (2000) NGC 5128 field, 
if viewed at the same distance and reddening as Maffei 
1, have $H = 21.3$, so the peak brightness in the Marleau et al. (2000) 
dataset is $\sim 0.6$ mag fainter in M$_H$ than in Maffei 1. However, this 
seeming difference in peak brightness is likely due to the relatively low 
stellar density in the NGC 5128 field, compounded 
by the modest angular coverage of NIC2. In fact, 
if the stellar contents in the Marleau et al. and Maffei 1 
fields are identical, then scaling the deep field star counts to the projected 
density of the NGC 5128 field indicates that only $0.3 \pm 0.2$ stars between 
$H = 20.25$ and 20.75 would be present in the Marleau et al. (2000) dataset; 
hence, there is only a modest chance of detecting stars within 0.5 mag of 
the AGB-tip with a single NIC2 pointing at this distance from the center 
of NGC 5128.

\section{THE CENTRAL REGIONS OF MAFFEI 1}

	Elliptical galaxies contain radial metallicity gradients (e.g. Davies, 
Sadler, \& Peletier 1993; Davidge 1997), in the sense 
that mean metallicity drops with increasing radius. 
The $V-I$ color of Maffei 1 becomes bluer towards larger 
radii (Buta \& McCall 1999), as expected if a metallicity gradient like that in 
other ellipticals is present. There is also a tendency 
for early-type galaxies in low-density environments to 
contain a centrally-concentrated component that is younger than the main 
body of the galaxy (e.g. Trager et al. 2000a). However, the tendency for $V-I$ 
to become redder with decreasing radius in Maffei 1 
is not consistent with such a component being present. 
Moreover, Luppino \& Tonry (1993) found that the characteristic 
flucuation brightness increases with radius in Maffei 1, although a similar 
gradient was not detected in M32 or the bulge of M31. 
Models by Blakeslee, Vazdekis, \& Ajhar (2001) and Liu, Graham, \& Charlot 
(2002) predict that this trend is in the opposite sense of what would be 
expected if a centrally-concentrated young population was present.

	Is there evidence for a centrally-concentrated young component 
in the central field data? To answer this question, the CFHT 
data were analyzed with the ellipse-fitting STSDAS task 
ELLIPSE after the $J$ and $H$ images were 
smoothed with a Gaussian to match the angular resolution of the $K$ data.
The $K-$band surface brightness profile, 
plotted in the upper panel of Figure 8, follows an r$^{1/4}$ law 
when log(r) $> 0$, in agreement with the profile measured by Buta 
\& McCall (1999) at shorter wavelengths and larger radii. 

	The $J-K$ color profile, shown in the lower panel of Figure 8, 
indicates that there is a modest red nucleus, which is confined to 
the central $1\arcsec$ of the galaxy.
The presence of a red nucleus is consistent with the 
$V-I$ color trend defined at much larger radii, although the apparent break in 
the $J-K$ profile at $r = 1\arcsec$ suggests that the central red component is 
distinct from the surrounding populations. If there were a central young 
population then one would expect to see a blue nucleus, although a large AGB 
component could cause an intermediate-age population to appear very red.

	The $J-K$ color curve in Figure 8 is qualitatively similar to that 
measured by Davidge \& Courteau (1999b) in M81, which contains a low-level AGN. 
However, there is no other evidence for an AGN in Maffei 1. 
Reynolds et al. (1997) detect only extended 
hard x-ray emission from Maffei 1, which they suggest originates from a 
population of low-mass x-ray binaries. If Maffei 1 contains an 
AGN then it must have a very modest energy output at x-ray wavelengths. 

\section{A SEARCH FOR BRIGHT GLOBULAR CLUSTERS}

	Maffei 1 is expected to host a substantial globular cluster population. 
With an integrated brightness of $V = 11.1$ (Buta \& McCall 1999) and 
A$_V = 4.5$ mag and $\mu_0 = 28.2$, then M$_V = -21.6$. 
If the specific frequency of globular clusters 
in Maffei 1 is similar to that in ellipticals in small 
groups, where $<S_N> = 2.6 \pm 0.5$ (Harris 1991), then 
the Maffei 1 cluster system should consist of $\sim 1130$ objects.

	The peak of the globular cluster LF (GCLF) in M31 occurs near $K_0 = 
14.5$ (Barmby, Huchra, \& Brodie 2001), which corresponds to M$_K = -10$. 
Therefore, adopting the M31 GCLF as a model, the peak in the Maffei 1 GCLF 
should occur near $K \sim 18.5$. There are a number of objects with 
$K \leq 18.5$ in the central field, and these are point sources in the 
$0\farcs35$ FWHM resolution data. The $(K, H-K)$ and $(K, J-K)$ CMDs of the 
bright objects in the central field are shown in Figure 9, along with the $(K, 
H-K)$ CMD of the DvdB background field. The sources near the center of Maffei 1 
have $H-K$ colors that match those of objects having similar brightness in the 
background field. In addition, the sources near the center of Maffei 1 
have $J-K < 1.1$, so that $(J-K)_0 < 0.3$; for comparison, Galactic globular 
clusters have $(J-K)_0 < 0.6$ (Brodie \& Huchra 1990). 
Hence, the bright objects in the deep field have $(J-K)_0$ 
colors that are not consistent with them being old globular clusters. 
While young, blue globular clusters have been detected near the centers of 
actively star-forming ellipticals (e.g. Carlson et al. 1998), there is no 
evidence for recent star formation near the center of Maffei 1, 
so it is unlikely that a population of young clusters would be 
present. We thus conclude that the central $700 \times 700$ 
parsec of Maffei 1 is devoid of clusters brighter than the peak of the GCLF. 

	Could some of the bright sources in the deep field be globular 
clusters? Lacking $J-K$ colors for objects in this field, we investigate 
this issue using statistical arguements. There are 5 sources with $H-K \sim 
0.2$ and $K < 19.5$ in the deep field. None of these are extended, and they 
have a density $\rho = 5/(14.4 \times 14.6) = 0.024$ arcsec $^{-2}$. 
Sources with similar brightness in the DvdB 
Maffei 1 and background fields have densities $\rho = 24/(34 \times 34) = 
0.021$ arcsec$^{-2}$, and $\rho = 12/(34 \times 34) = 0.010$ arcsec $^{-2}$. 
The mean density of objects with $K < 19.5$ in all three 
fields, weighted according to field area, is then 
$\overline{\rho} = 0.016$ arcsec$^{-2}$, indicating that 
there should be $\sim 3.3$ sources with $K \leq 19.5$ in the deep field if 
objects of this brightness are uniformly distributed. The Poisson probability 
function then indicates that there is only a 12\% chance that 5 objects 
with $K \leq 19.5$ would be detected in the deep field. These data thus hint 
that the deep field may contain an excess of objects with $K \leq 19.5$ when 
compared with larger radii.

\section{DISCUSSION \& SUMMARY}

\subsection{The Stellar Content of Maffei 1 and Other Spheroids}

	Deep $H$ and $K'$ images obtained with the UHAO $+$ QUIRC 
on the Northern Gemini telescope have been used to investigate the bright 
AGB content of a field 3 arcmin ($\sim 4$ kpc) from the center of 
Maffei 1. If it is assumed that the brightest giants in Maffei 1 have 
the same intrinsic $H-K$ color as late M giants in BW then a line-of-sight 
extinction is computed that is consistent with previous estimates, which have 
relied largely on the integrated properties of the galaxy.

	The main result of this paper is that the infrared-bright stellar 
content of the Maffei 1 deep field, as gauged by (1) the brightness of the 
AGB-tip, (2) the shape of the AGB LF, and (3) the density of AGB stars measured 
with respect to surface brightnesses at visible wavelengths, does not differ 
significantly from that in other nearby spheroids. The AGB-tip in the 
Maffei 1 deep field occurs near M$_K \sim -8.7$, and thus is comparable to the 
peak brightnesses in M32 (Davidge 2000a; Davidge et al. 2000), and the bulges 
of the Milky-Way and M31 (Davidge 2001a). Rejkuba et al. 
(2001) find that the peak M$_K$ in the outer regions of NGC 5128 is $\sim 
-8.8$, which is also in remarkable agreement with the peak brightness in 
Maffei 1.

	The near-infrared LFs of bright stars in the Maffei 1 deep 
field, and the outer regions of M32 and NGC 5128 are in excellent 
agreement. In some respects, the good agreement 
between the bright stellar contents of Maffei 1 and NGC 5128 is 
perhaps not surprising, given that these galaxies have comparable 
integrated brightnesses, distances, and environments. However, 
the chemical enrichment history of a galaxy is thought to depend on 
factors such as galaxy mass (e.g. Yoshii \& Arimoto 1987), and it might be 
anticipated that the photometric properties of the brightest stars 
in a massive elliptical galaxy like Maffei 1 might 
differ from those in a smaller system like M32, due to differences in 
metallicity. Indeed, the integrated Mg$_2$ index of M32 is markedly lower 
than in more massive ellipticals (Burstein et al. 1984). 
However, the metallicity distribution function (MDF) of M32 
measured by Grillmair et al. (1996) is similar to that of the outer regions of 
NGC 5128 (Harris \& Harris 2000; Harris, Harris, \& Poole 1999), suggesting 
that the stellar contents of M32 and larger ellipticals may not be so different.

	Insight into the nature of the brightest stars in nearby spheroids 
can be obtained by examining their distribution within 
these systems. In M32 and the bulge of M31 the brightest stars 
are uniformly distributed throughout these systems, with a number density that 
scales with $r-$band surface brightness (Davidge 2000a, Davidge 
et al. 2000, and Davidge 2001). There are indications that the brightest 
stars also are uniformly distributed in NGC 5128, as Harris \& 
Harris (2000) show that the relative number density of the brightest 
AGB and RGB stars does not change with radius, although 
it is evident from their Figures 7 and 8 that their data are not 
sensitive to luminous giants with solar or higher metallicities. Curiously, 
a comparison of the $K$ LFs of the DvdB field and the deep field 
suggests that the outer regions of Maffei 1 may be deficient in stars near $K = 
20.5$, suggesting that the brightest stars in Maffei 1 may 
not be uniformly distributed throughout the entire galaxy. 
A survey of the outer regions of Maffei 1, covering a square arcmin or more and 
sampling stars as faint as $K = 21$ would provide the data that is needed to 
confirm if the deficiency of bright stars in the DvdB field is real, or a 
statistical fluke.

	Soria et al. (1996) detected stars as bright as M$_{bol} \sim -5$ 
in the inner halo of NGC 5128, and concluded that these objects belong to an 
intermediate-age population. Marleau et al. (2000) reached a similar conclusion 
after analyzing near-infrared observations of a portion of the Soria et al. 
field, and it is these data that have been compared with the Maffei 1 deep 
field observations. However, peak AGB luminosity is not an ironclad means 
of judging the age of a population, as the peak AGB brightness is a function of 
metallicity as well as age, and this introduces uncertainty in the age 
calibration of the AGB-tip. In fact, Guarnieri, Renzini, \& Ortolani 
(1997) examined the brightest members of moderately metal-rich globular 
clusters, which have old ages (e.g. Ortolani et al. 1995), and found that the 
brightest AGB stars have M$_{bol}$ between --4.5 and --5.0 when [Fe/H] $> 
-1.0$; thus, the bright stars detected by Soria et al. (1996) may have ages 
comparable to Galactic metal-rich globular clusters. 

	The galaxy-to-galaxy similarity in peak M$_K$ and steller density 
are difficult to explain if the brightest stars are young or of 
intermediate-age, as these systems must then have experienced fortuitously 
similar star-forming histories during intermediate epochs: not only would age 
and metallicity have to be tuned to produce similar peak AGB luminosities, but 
the intermediate-age components would also have to be uniformly distributed 
throughout these systems with similar spatial densities. Both of these 
difficulties vanish if the bright stars are old; in this case 
the problem of tuning the AGB-tip brightness is 
less acute because the rate of change of this parameter with time decreases 
with increasing age. Likewise, the uniform distribution of infrared-bright 
stars occurs naturally if they formed during the initial collapse of the 
system, when the main structural characteristics of the galaxies 
were defined and there was likely a system-wide period of star formation. 
Finally, stars with a peak brightness like that in Maffei 1, M32, and the 
bulge of M31 occur in the Galactic Bulge (Davidge 2001a), which appears to have 
an old age (Feltzing \& Gilmore 2000; Ortolani et al. 1995).

	Clearly, NGC 5128 has experienced recent star formation, 
with the younger populations being centrally concentrated. 
However, based on the comparison with Maffei 1, the main body of this galaxy 
is old. It thus appears that NGC 5128 is a nearby example of the 
`frosting' model proposed to explain the integrated spectroscopic 
properties of many early-type galaxies in the field,
in which a modest young or intermediate age population is 
superimposed on an old stellar substrate (Trager et al. 2000b).

	Hierarchal models of galaxy formation, which assume that large 
galaxies are assembled by the accretion of smaller systems, 
are able to reproduce many observed properties of present-day galaxies (e.g. 
Somerville \& Primack 1999; Cole et al. 2000). 
One prediction of these models is that 50\% of all stars formed prior to 
z = 1.5 (Cole et al. 1994; 2000). It can be anticipated that 
most of the stars (or their remnants) that formed prior to z = 1.5 will 
be in spheroidal systems at the current epoch, since 
mergers and feedback from star formation likely 
prevented disks from forming until z $\sim 1$ (e.g. 
Weil, Eke, \& Efstathiou 1998). That spheroids are dominated by stars that 
formed early-on is consistent with the Mg$_2 - \sigma_0$ relation of these 
systems, which indicates that their basic structural properties 
were imprinted at high redshift (Bernardi et al. 1998).

	It is somewhat suprising that the brightest stars appear 
to be uniformly distributed throughout the main bodies of systems like M32, 
the bulge of M31, and NGC 5128, as the evolution of a region within a galaxy is 
influenced by the local mass density, which defines the escape velocity and 
(possibly) the star formation rate (e.g. Schmidt 1959). A radial variation in 
escape velocity may be the physical basis for metallicity gradients in 
elliptical galaxies (Franx \& Illingworth 1991; Martinelli, Matteucci, \& 
Colafrancesco 1998), as well as the tight relations 
between absorption line strengths and local surface 
brightness (Kobayashi \& Arimoto 1999; Davidge \& Grinder 
1995). Local surface brightness is a relative measure of mass density, at least 
to the extent that spheroidal systems have similar M/L ratios. The surface 
brightnesses of the various fields that have been compared in this paper and in 
Davidge (2001a) are summarized in Table 1, and it is evident that these 
span a wide range of values. These data ostensibly suggest 
that the progenitors of the bright AGB stars studied in this paper can form in 
regions with surface brightnesses in M$_V$ at least as low as $\sim 1$ mag 
pc$^{-2}$ ($\sim 30$ M$_{\odot}$ pc$^{-2}$).

\subsection{The Central Regions of Maffei 1}

	The data presented in this paper indicate that Maffei 1 contains a red 
nucleus, that extends out to $\sim 1\arcsec$. The nature of this nucleus is not 
clear, although there are hints that it is not a low level AGN. The 
absence of a discrete x-ray point source in Maffei 
1 has been noted by Reynolds et al. (1997). In addition, 
Spinrad et al. (1971) discuss the only spectroscopic observations 
of Maffei 1 that are known to us. Their spectrum, obtained with a $2\arcsec$ 
wide slit, shows strong line absorption, with no hint of central line or 
continuum emission.

	While there is no evidence for a systematic age gradient in Maffei 
1 (\S 5), the presence of a young nucleus can not be completely discounted. 
However, if the nucleus is younger than the main body of the galaxy then 
it must be heavily extincted and/or viewed at an evolutionary stage when the 
AGB dominates the infrared light. Buta \& McCall 
(1999) do find dust north of the nucleus of Maffei 1.

	As a moderately large (M$_V \sim -21.6$) elliptical galaxy, Maffei 1 
should have a well-populated globular cluster system. However, the central $700 
\times 700$ parsecs of Maffei 1 is devoid of globular clusters brighter than 
the peak of the GCLF. While dynamical evolution is expected to 
disrupt clusters in the central regions of galaxies (e.g. Portegies 
Zwart et al. 2001; Murali \& Weinberg 1997; Vesperini 1997), some nearby 
elliptical galaxies have bright globular clusters within 
a few hundred parsecs of their centers. Forbes et al. (1996) 
investigated the central globular cluster contents of a sample of elliptical 
galaxies. One of the nearest galaxies in their sample is NGC 4494, 
which has an integrated brightness similar to Maffei 1; 6 clusters brighter 
than the peak of the GCLF were found within 400 parsecs of the center of 
this galaxy. Interestingly, despite having what appears to be a well-populated 
central cluster system, NGC 4494 may have a lower than average 
global specific cluster frequency (Larsen et al. 2001).

	The specific globular cluster frequency of Maffei 1 is not known. 
Because of the heavy extinction at visible wavelengths any 
survey for globular clusters in Maffei 1 should likely be conducted in the 
near-infrared. Based on the relation between the globular 
cluster system core radius and host galaxy brightness calibrated by Forbes et 
al. (1996), the core radius of the Maffei 1 cluster system should be $\sim 2.5$ 
kpc, so a number of clusters should be present within $\sim 2$ arcmin of the 
galaxy center. Foreground star contamination is an obvious concern, although 
this does not present an insurmountable hurdle, 
since field stars with brightnesses comparable to 
those of bright clusters in Maffei 1 have relatively blue colors 
(\S 6). Hence, it should be possible to distinguish between clusters and stars 
using $J-K$ colors.

\acknowledgments

It is a pleasure to thank Kathy Roth, Olivier Guyon, Pierre Baudoz, and Brian 
Walls for acting as my hands and eyes at the Gemini telescope during a time 
when travel to Hawaii was not possible. Sincere thanks are also extended 
to Sidney van den Bergh for commenting on an earlier draft of this 
manuscript, and to an anonymous referee, whose comments helped to clarify 
many of the points discussed in this paper.

\parindent=0.0cm

\clearpage

\begin{table*}
\begin{center}
\begin{tabular}{lcl}
\tableline\tableline
Field & $V$ Surface Brightness & Source \\
 & (mag pc$^{-2}$) & \\
\tableline
M31 Bulge Field & --2.8 & Davidge (2001a) \\
Maffei 1 Deep Field & --1.1 & This paper \\
Maffei 1 DvdB Field & 0.3 & DvdB \\
M32 Outer Field & 1.0 & Davidge (2000a) \\
NGC 5128 NIC2 Field & 1.2 & Marleau et al. (2000) \\
\tableline
\end{tabular}
\end{center}
\caption{Surface brightness measurements of fields in spheroidal systems where upper AGB stars have been resolved}
\end{table*}

\clearpage
\begin{center}
FIGURE CAPTIONS
\end{center}

\figcaption
[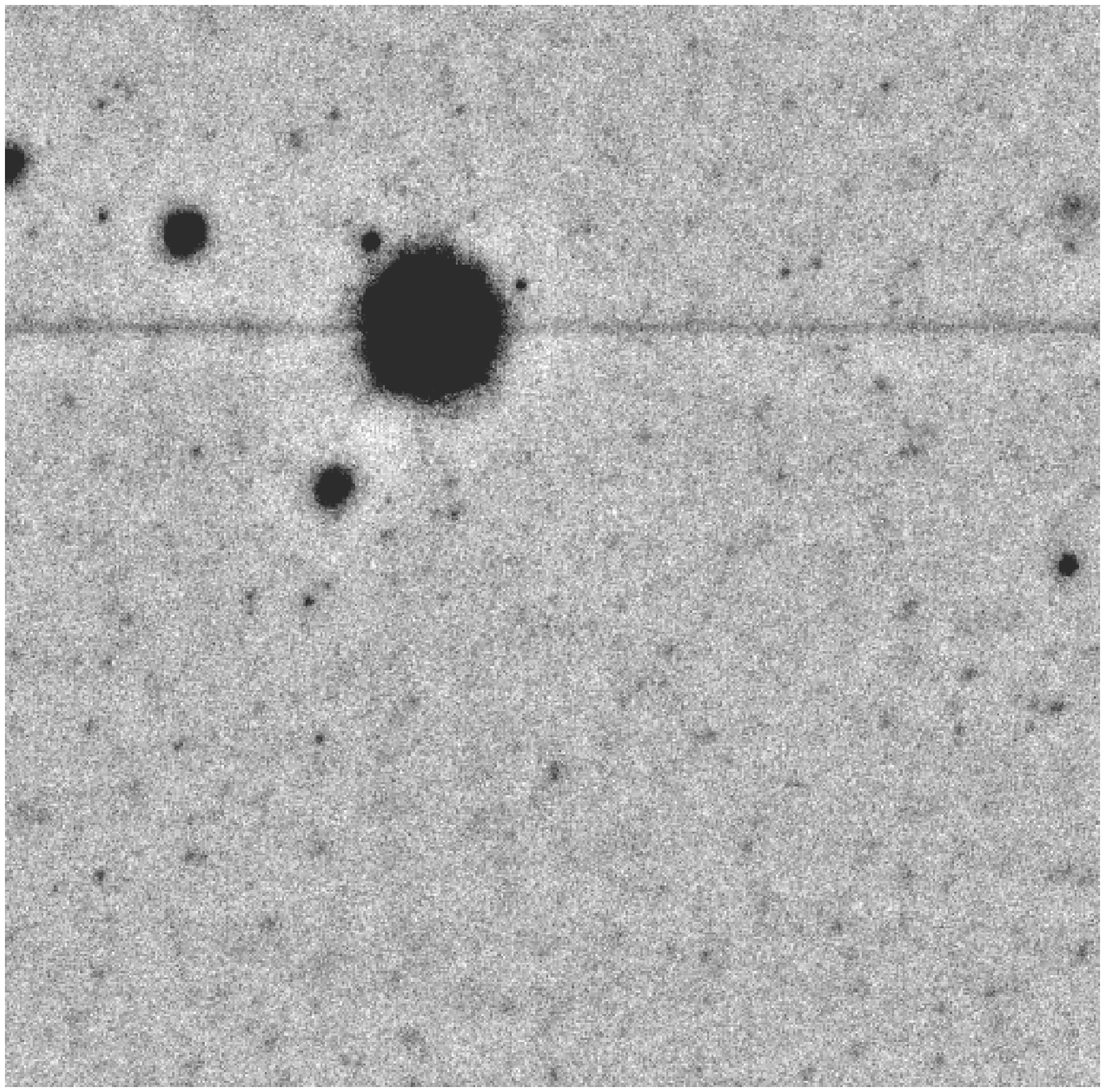]
{The final $K'$ image of the deep field, which covers $14\farcs6 \times 
14\farcs4$. North is at the top, and East is to the left. The angular 
resolution is $0\farcs14$ FWHM. The bright object to the upper left of 
the field center is the $R = 13.4$ star GSC 03699-01165, which was the 
reference source for AO correction. The faint sources scattered throughout the 
field are stars in Maffei 1.}

\figcaption
[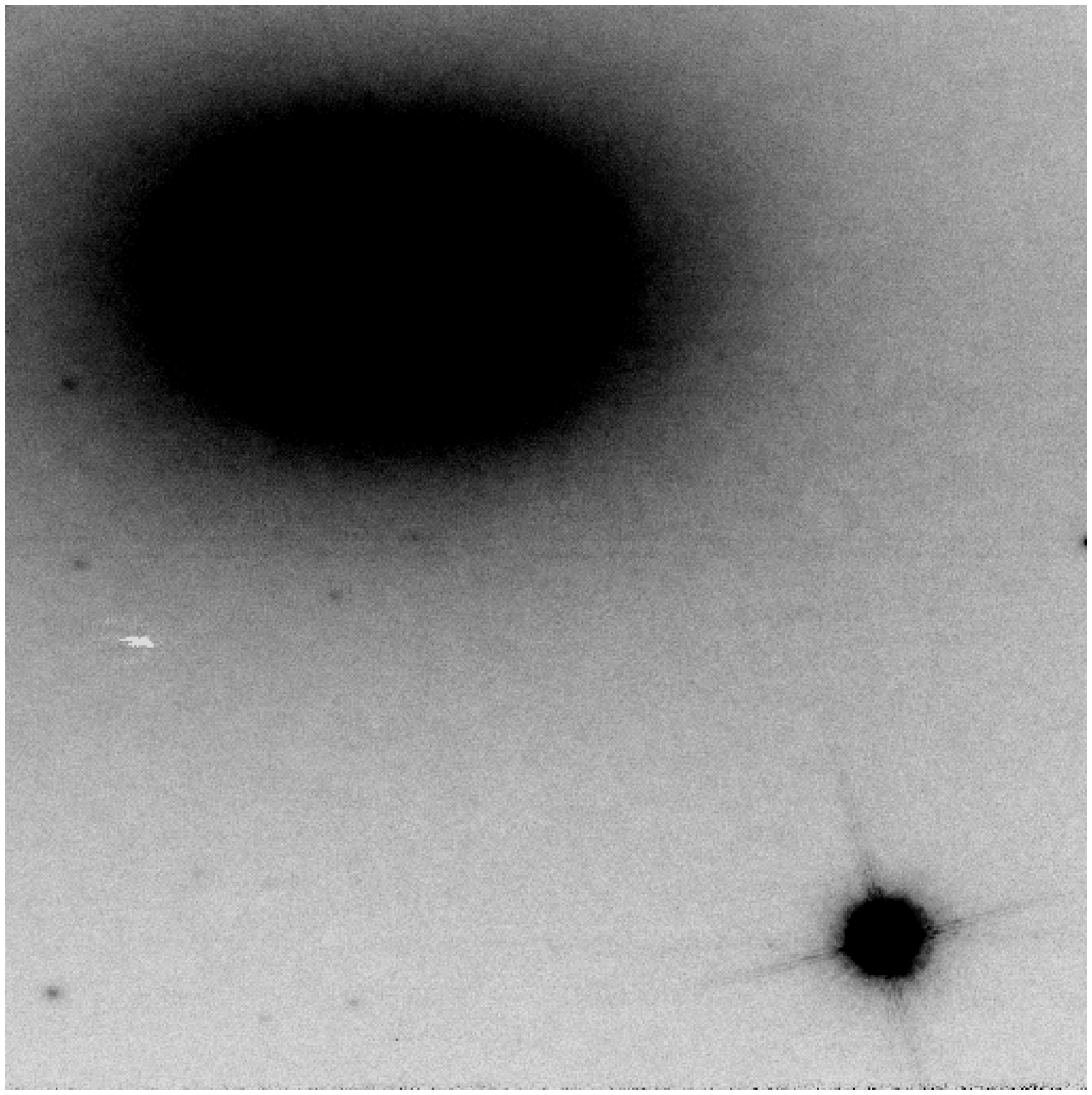]
{The final $Ks$ image of the central field, which covers $34\farcs0 \times 
34\farcs0$ arcsec. North is at the top, and East is to the left. The angular 
resolution is $0\farcs35$ FWHM. The bright source in the lower right hand 
corner is the $R = 12.3$ star GSC 03699-01147, which was the reference 
source for AO correction. The several faint point sources scattered throughout 
the field have $K > 18$, and are likely foreground disk stars (\S 6).}

\figcaption
[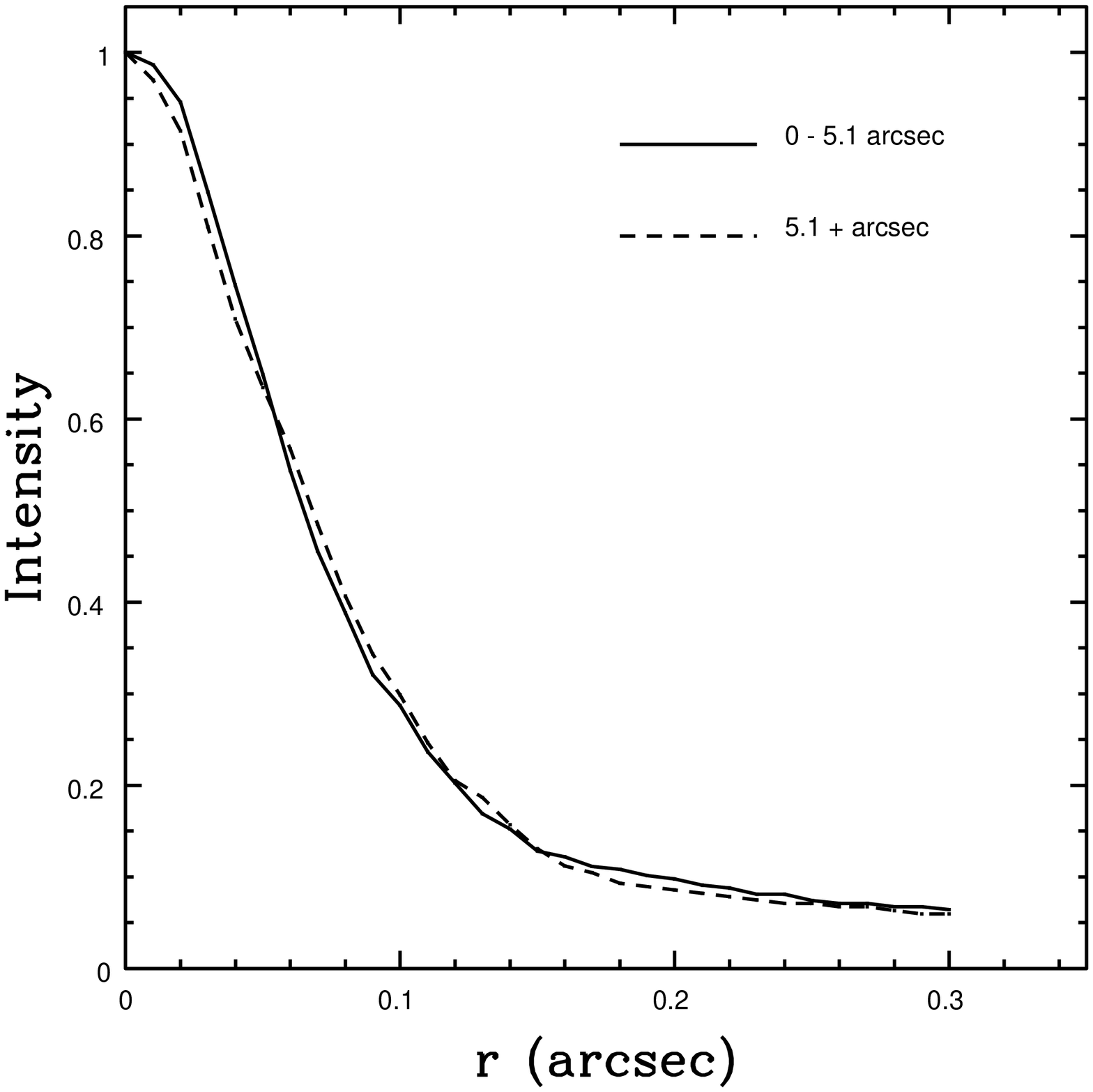]
{The mean $K'$ light profiles of stars in two radial intervals in the Maffei 1 
deep field. The solid line shows the PSF for stars within $5\farcs1$ of 
the AO guide star, while the dashed line shows the PSF for stars with $r \geq 
5\farcs1$; these radial intervals sample comparable areas in the Maffei 1 deep 
field. The curves were constructed from stars having similar brightnesses 
so as not to bias the comparison towards a particular radius. Note the 
excellent agreement between the two curves, indicating that the PSF 
is stable across the field.}

\figcaption
[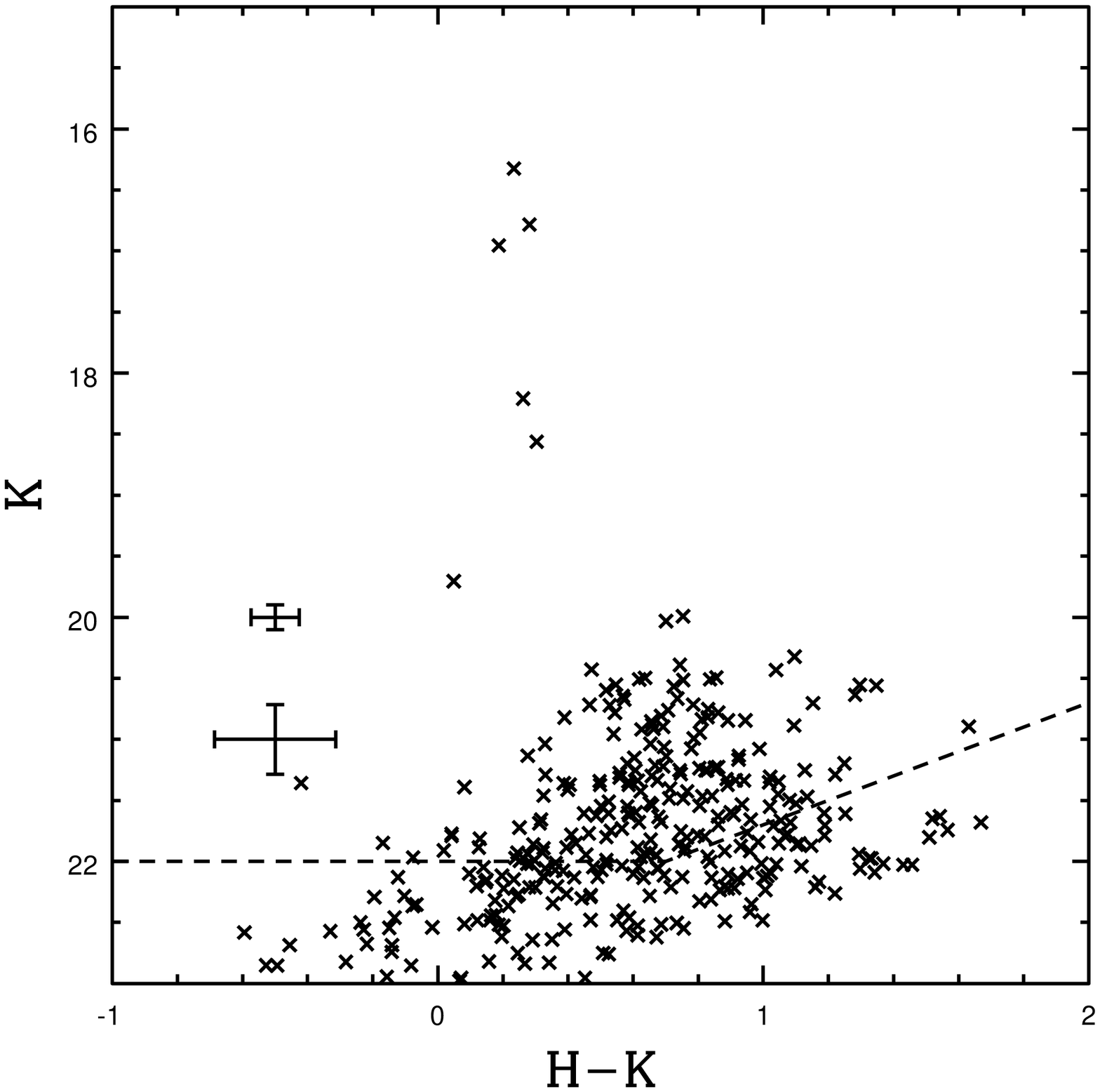]
{The $(K, H-K)$ CMD of the Maffei 1 deep field. Stars in Maffei 1 
form the broad plume with $K < 20$ centered near $H-K = 0.7$. The 
majority of sources with $K < 19.5$ are 
likely foreground stars in the Galactic disk. The error bars show the 
1$-\sigma$ uncertainties predicted from the artificial star experiments, while 
the dashed line shows the 50\% completeness level.}

\figcaption
[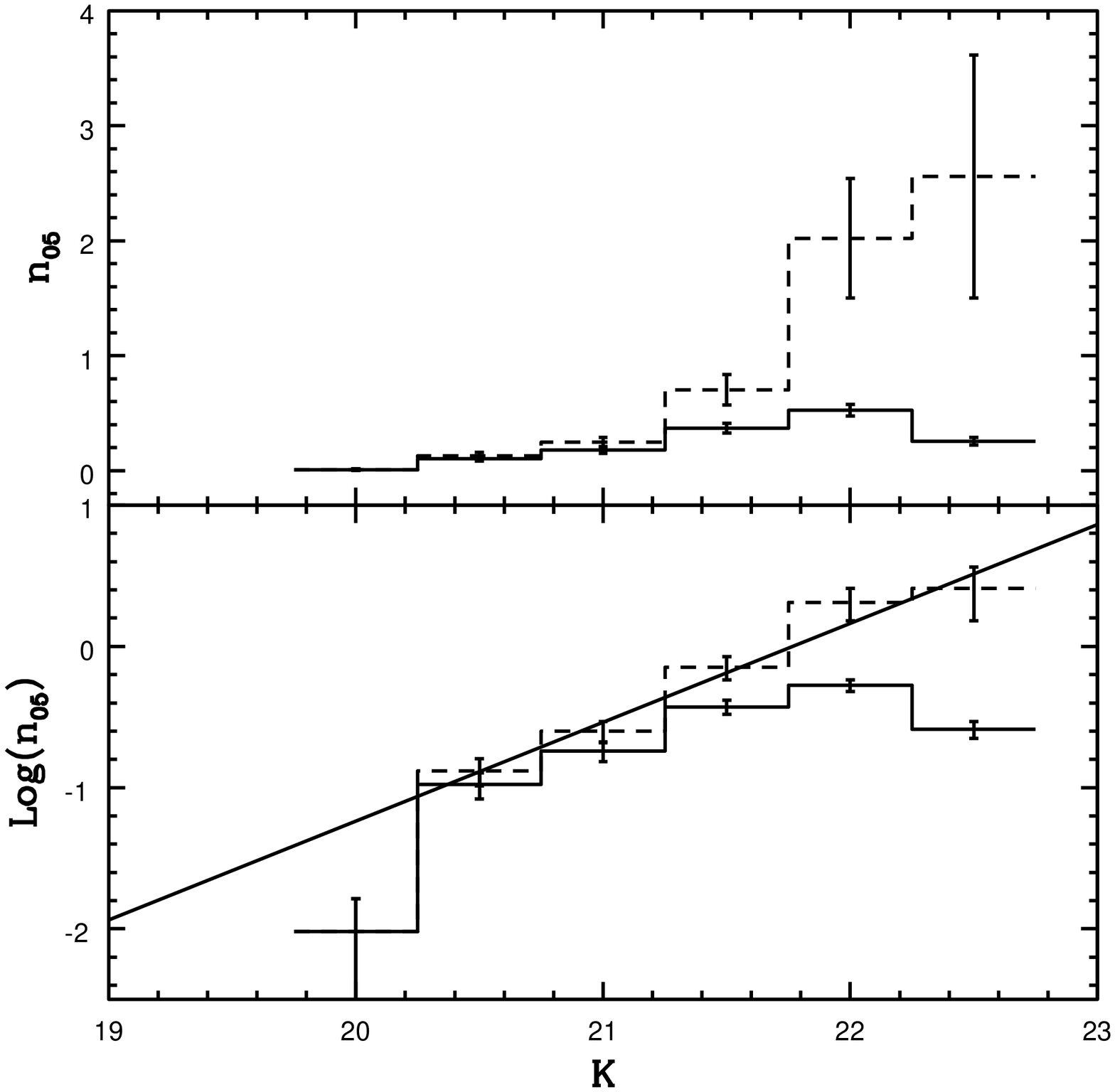]
{The $K$ LF of the Maffei 1 deep field. n$_{0.5}$ is the number of stars 
per square arcsec per 0.5 mag interval in $K$. The solid line shows the raw LF 
constructed from stars detected in both $H$ and $K'$, while 
the dashed line is the LF corrected for incompleteness. The error bars include 
counting statistics and the uncertainities in the completeness corrections. 
Also shown in the lower panel is a power-law fit to the 
completeness-corrected LF when $K > 20.25$, which has an exponent 
$x = 0.70 \pm 0.07$.}

\figcaption
[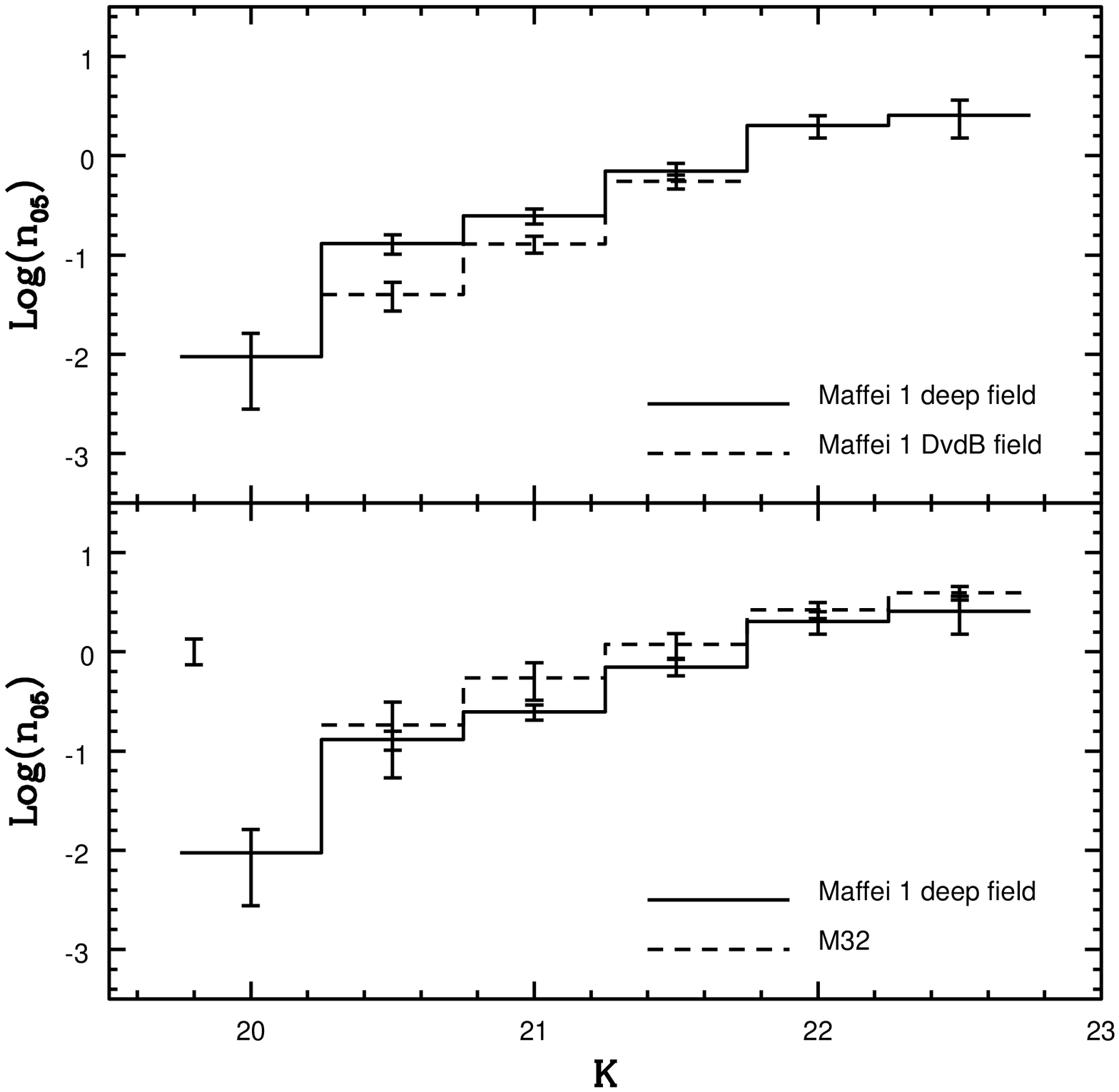]
{The $K$ LF of the Maffei 1 deep field compared with the $K$ LF 
of the field studied by DvdB (upper panel) and the M32 outer field observed 
by Davidge (2000a) (lower panel). The LFs in this figure have been corrected 
for incompleteness, and the errorbars show the uncertainties due to counting 
statistics and the completeness corrections. n$_{0.5}$ is the number of stars 
per 0.5 mag per square arcsec in the deep field. The M32 and Maffei 1 DvdB 
LFs were shifted along the vertical axis to match the stellar density in 
the Maffei 1 deep field using the published surface brightness profiles 
given in the text, while the brightnesses of stars in M32 have been shifted 
to match the distance and reddening of Maffei 1. 
The errorbar in the upper left hand corner of the lower panel 
shows the combined systematic uncertainty in the relative distances 
and reddenings of M32 and Maffei 1; note that this error is comparable 
to the offset between the M32 and Maffei 1 LFs.}

\figcaption
[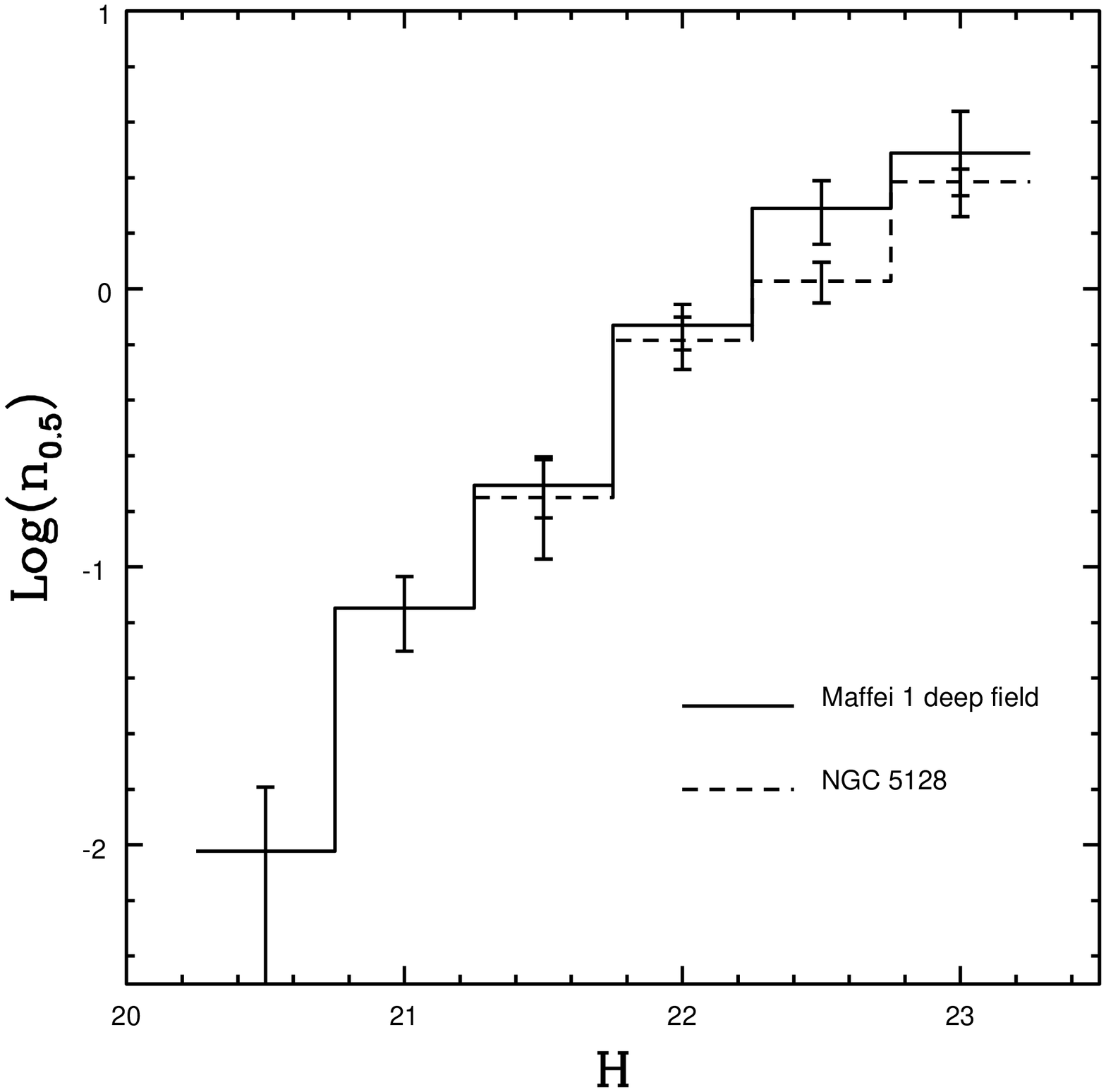]
{The $H$ LFs of the Maffei 1 deep field (solid line) and 
the Marleau et al. (2000) NGC 5128 field (dashed line). 
The brightnesses of stars in NGC 5128 were 
shifted to match the distance and reddening of Maffei 1 prior to 
constructing the LF for this galaxy, and the result was then scaled 
along the vertical axis to match the stellar density in the Maffei 1 field 
using published surface brightness measurements. The errorbars 
show the uncertainties due to counting statistics and the completeness 
corrections; n$_{0.5}$ is the number of stars per 0.5 mag interval 
in $H$ per square arcsec in the Maffei 1 deep field.
Note the excellent agreement between the LFs of these galaxies.} 

\figcaption
[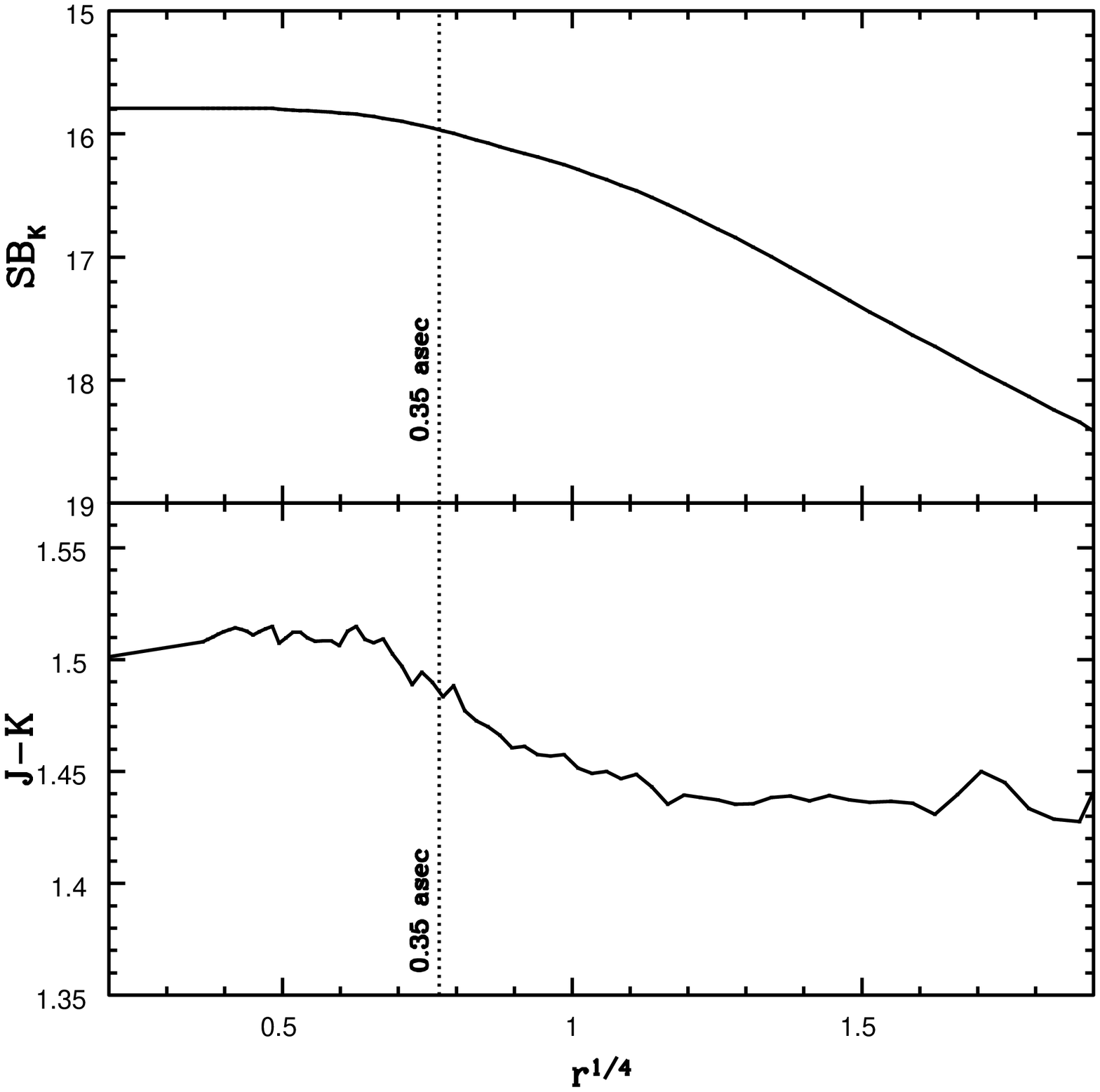]
{The $K-$band surface brightness and $J-K$ color profiles near the center 
of Maffei 1. $r$ is the distance in arcsec from the galaxy center, 
and the dotted line indicates 0.35 arcsec, which is the angular 
resolution of these data. Note the appearance of a red nucleus when $r < 
1\arcsec$.}

\figcaption
[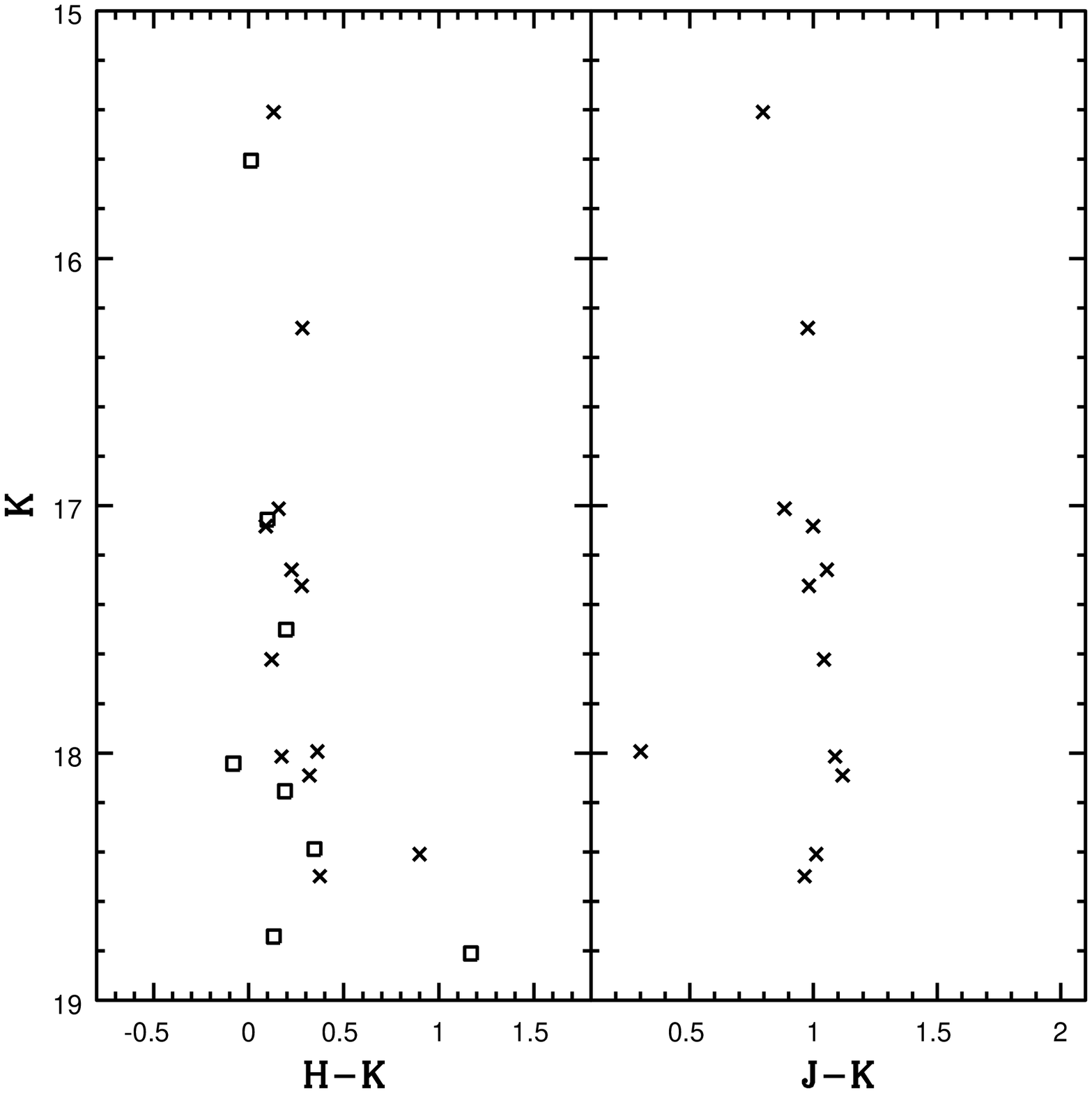]
{The $(K, H-K)$ and $(K, J-K)$ CMDs of sources with $K < 18.5$ in the 
Maffei 1 central field. Sources in the DvdB background field are plotted as 
open squares in the $(K, H-K)$ CMD. Note the excellent agreement 
between the central and background field datapoints on the $(K, H-K)$ CMD.}


\clearpage

\begin{references}

\reference{}Barmby, P., Huchra, J. P., Brodie, J. P. 2001, AJ, 121, 1482

\reference{}Bernardi, M. et al. 1998, ApJ, 508, L43

\reference{}Bertelli, G., Bressan, A., Chiosi, C., Fagotto, F., \& Nasi, E. 1994, A\&AS, 106, 275

\reference{}Blakeslee, J. P., Vazdekis, A., \& Ajhar, E. A. 2001, MNRAS, 320, 193

\reference{}Brodie, J. P., \& Huchra, J. P. 1990, ApJ, 362, 503

\reference{}Burstein, D., Faber, S. M., Gaskell, C. M., \& Krumm, N. 1984, ApJ, 287, 586

\reference{}Buta, R. J., \& McCall, M. L. 1983, MNRAS, 205, 131

\reference{}Buta, R. J., \& McCall, M. L. 1999, ApJ, 124, 33

\reference{}Carlson, M. N. et al. 1998, AJ, 115, 1778

\reference{}Cole, S., Arag\'{o}n-Salamanca, A., Frenk, C. S., Navarro, J. F., \& Zepf, S. E. 1994, MNRAS, 271, 781

\reference{}Cole, S., Lacey, C. G., Baugh, C. M., \& Frenk, C. S. 2000, MNRAS, 319, 168

\reference{}Davidge, T. J. 1991, ApJ, 380, 116

\reference{}Davidge, T. J. 1997, AJ, 113, 985

\reference{}Davidge, T. J. 2000a, PASP, 112, 1177

\reference{}Davidge, T. J. 2000b, AJ, 120, 1853

\reference{}Davidge, T. J. 2001a, AJ, 122, 1386

\reference{}Davidge, T. J. 2001b, AJ, 121, 3100

\reference{}Davidge, T. J., \& Courteau, S. 1999a, AJ, 117, 1297

\reference{}Davidge, T. J., \& Courteau, S. 1999b, AJ, 117, 2781

\reference{}Davidge, T. J., \& Courteau, S. 2002, AJ, 123, 1438

\reference{}Davidge, T. J., \& Grinder, M. 1995, AJ, 109, 1433

\reference{}Davidge, T. J., \& van den Bergh, S. 2001, ApJ, 553, L133

\reference{}Davidge, T. J., Rigaut, F., Chun, M., Brandner, W., Potter, D., Northcott, M., \& Graves, J. E. 2000, ApJ, 545, L89

\reference{}Davies, R. L., Sadler, E. M., \& Peletier, R. 1993, MNRAS, 262, 650

\reference{}DePoy, D. L., Terndrup, D. M., Frogel, J. A., Atwood, B., \& Blum, R. 1993, AJ, 105, 2121

\reference{}Feltzing, S., \& Gilmore, G. 2000, A\&A, 355, 949

\reference{}Forbes, D. A., Franx, M., Illingworth, G. D., \& Carollo, C. M. 1996, ApJ, 467, 126

\reference{}Franx, M., \& Illingworth, G. 1991, ApJ, 359, L41

\reference{}Frei, Z., \& Gunn, J. E. 1994, AJ, 108, 1476

\reference{}Frogel, J. A., \& Whitford, A. E. 1987, ApJ, 320, 199

\reference{}Graves, J. E., Northcott, M., Roddier, F., Roddier, C., \& Close, L. 1998, Proc. SPIE 3353, 34

\reference{}Grillmair, C. J. et al. 1996, AJ, 112, 1975

\reference{}Guarnieri, M. D., Renzini, A., \& Ortolani, S. 1997, ApJ, 477, L21

\reference{}Harris, W. E. 1991, ARA\&A, 29, 543

\reference{}Harris, G. L. H., \& Harris, W. E. 2000, AJ, 120, 2423

\reference{}Harris, G. L. H., Harris, W. E., \& Poole, G. B. 1999, AJ, 117, 855

\reference{}Hawarden, T. G., Leggett, S. K., Letawsky, M. B., Ballantyne, D. R., \& Casali, M. M. 2001, MNRAS, 325, 563

\reference{}Israel, F. P. 1998, A\&AR, 8, 237

\reference{}Kent, S. M. 1987, AJ, 94, 306

\reference{}Kobayashi, C., \& Arimoto, N. 1999, ApJ, 527, 573

\reference{}Larsen, S. S., Brodie, J. P., Huchra, J. P., Forbes, D. A., \& Grillmair, C. J. 2001, AJ, 121, 2974

\reference{}Liu, M. C., Graham, J. R., \& Charlot, S. 2002, ApJ, 564, 216

\reference{}Luppino, G. A., \& Tonry, J. L. 1993, ApJ, 410, 81

\reference{}Marleau, F. R., Graham, J. R., Liu, M. C., \& Charlot, S. 2000, AJ, 120, 1779

\reference{}Martinelli, A., Matteucci, F., \& Colafrancesco, S. 1998, MNRAS, 298, 42

\reference{}McClure, R. D., Arnaud, J., Fletcher, J. M., Nieto, J-L, \& Racine, R. 1991, PASP, 103, 570

\reference{}Murali, C., \& Weinberg, M. D. 1997, MNRAS, 288, 749

\reference{}Ortolani, S. et al. 1995, Nature, 377, 701

\reference{}Portegies Zwart, S. F., Makino, J., McMillan, S. L. W., \& Hut, P. 2001, ApJ, 546, L101

\reference{}Rejkuba, M., Minniti, D., Silva, D. R., \& Bedding, T. R. 2001, A\&A, 379, 781

\reference{}Reynolds, C. S., Loan, A. J., Fabian, A. C., Makishima, K., Brandt, W. N., \& Mizuno, T. 1997, MNRAS, 286, 349

\reference{}Rieke, G. H., \& Lebofsky, M. J. 1985, ApJ, 288, 618

\reference{}Rigaut, F. et al. 1998, PASP, 110, 152

\reference{}Schmidt, M. 1959, ApJ, 129, 243

\reference{}Somerville, R. S., \& Primack, J. R. 2000, MNRAS, 310, 1087

\reference{}Soria, R., et al. 1996, ApJ, 465, 79

\reference{}Spinrad, H. et al. 1971, ApJ, 163, L25

\reference{}Stetson, P. B. 1987, PASP, 99, 191

\reference{}Stetson, P. B., \& Harris, W. E. 1988, AJ, 96, 909

\reference{}Trager, S. C., Faber, S. M., Worthey, G., \& Gonzalez, J. J. 2000a, AJ, 119, 1645

\reference{}Trager, S. C., Faber, S. M., Worthey, G., \& Gonzalez, J. J. 2000b, AJ, 120, 165

\reference{}van den Bergh, S. 1976, ApJ, 208, 673

\reference{}van den Bergh, S. 2000, The Galaxies of the Local Group (Cambridge: Cambridge University Press), 11

\reference{}Vesperini, E. 1997, MNRAS, 287, 915

\reference{}Wainscoat, R. J., \& Cowie, L. L. 1992, AJ, 103, 332

\reference{}Weil, M. L., Eke, V. R., \& Efstathiou, G. 1998, MNRAS, 300, 773

\reference{}Yoshii, Y., \& Arimoto, N. 1987, A\&A, 188, 13

\end{references}
\end{document}